\documentclass[twocolumn,amsmath,amssymb,aps,pra,floatfix,nofootinbib,superscriptaddress,showpacs]{revtex4}
\usepackage{bm}
\usepackage{longtable}
\usepackage{dcolumn}
\usepackage{bm}
\usepackage{amsthm}
\newtheorem{theorem}{Theorem}
\newtheorem{example}{Example}
\usepackage[pdftex]{graphicx}
\graphicspath{{figures/pdf/}} \DeclareGraphicsExtensions{.pdf}
\usepackage{times,color}

\def\ket#1{| #1 \rangle}
\def\bra#1{\langle #1 |}

\def\ave#1{\langle #1 \rangle}

\def\RR{\mathbb{R}}
\def\diag{\operatorname{diag}}

\def\dim{\operatorname{dim}}

\def\Span{\operatorname{span}}

\def\Tr{\operatorname{Tr}}

\def\H{\mathcal{H}}

\def\L{\mathcal{L}}
\def\RR{\mathcal{R}}

\def\uu{\mathbf{u}}
\def\su{\mathbf{su}}
\def\SU{\mathbb{SU}}
\def\UU{\mathbb{U}}

\def\la{\langle}
\def\ra{\rangle}

\begin{document}
\title{Robust Entanglement in Anti-ferromagnetic Heisenberg Chains by
Single-spin Optimal Control}

\author{Xiaoting Wang} \email{xw233@cam.ac.uk}
\affiliation{Department of Applied Maths and Theoretical Physics,
University of Cambridge, Wilberforce Rd, Cambridge, CB3 0WA, United
Kingdom}

\author{Abolfazl Bayat} \email{abolfazl.bayat@ucl.ac.uk}
\affiliation{Department of Physics and Astronomy, University College
London, Gower St., London WC1E 6BT, United Kingdom}

\author{S.~G.~Schirmer}\email{sgs29@cam.ac.uk}
\affiliation{Department of Applied Maths and Theoretical Physics,
University of Cambridge, Wilberforce Rd, Cambridge, CB3 0WA, United
Kingdom}

\author{Sougato Bose} \email{sougato@theory.phys.ucl.ac.uk}
\affiliation{Department of Physics and Astronomy, University College
London, Gower St., London WC1E 6BT, United Kingdom}

\date{\today}
\begin{abstract}
We demonstrate how near-perfect entanglement (in fact arbitrarily close
to maximal entanglement) can be generated between the end spins of an
anti-ferromagnetic isotropic Heisenberg chain of length $N$, starting
from the ground state in the $N/2$ excitation subspace, by applying a
magnetic field along a given direction, acting on a single spin only.
Temporally optimal magnetic fields to generate a singlet pair between
the two end spins of the chain are calculated for chains up to length 20
using optimal control theory.  The optimal fields are shown to remain
effective in various non-ideal situations including thermal
fluctuations, magnetic field leakage, random system couplings and
decoherence. Furthermore, the quality of the entanglement generated can
be substantially improved by taking these imperfections into account in
the optimization. In particular, the optimal pulse of a given thermal
initial state is also optimal for any other initial thermal state with
lower temperature.

\end{abstract}

\maketitle

\section{Introduction}

Spin systems are important models to understand many-body physics
and to investigate many interesting problems in the area of quantum
information processing~\cite{Amico-RMP}.  Coherent control of such
systems to achieve certain key tasks such as entanglement generation
is one of the main problems in quantum mechanics. There have been
several proposals already for entanglement generation between the
end points of a spin chain both by distributing the entanglement
through the chain \cite{bose,bayat-xxz} and entanglement generation
through a quench followed by the natural non-equilibrium dynamics of
the system \cite{bayat-quench,hannu-quench,hangi-quench}.
Unfortunately, the attainable entanglement in these schemes is not
maximal and achieving an arbitrarily high entanglement between
ending spins demands more control.  Although there has been
significant progress in both the theory and implementation of
coherent control, the microscopic nature of most quantum objects
makes it difficult to control all elements of the system
individually.  Local addressing of individual spins, for example,
remains a significant challenge. Applying fields that globally
address the entire system is one way to avoid such
problems~\cite{fitzsimon}. On the other hand, if assuming locally
addressing ability of a many-body system, one interesting question
is what is the minimal controlling resource that can perform certain
desired task. For example, universal quantum computation can be
achieved through controlling a small part or gateway of the whole
system~\cite{Lloyd,Burgarth-Bose, PRA80n030301(R), Burgarth,Kay},
when the system intrinsic Hamiltonian provides the necessary
non-local interactions for a successful control. Aside from not
requiring access to the entire system, limiting the interaction of
the controller with the system can be advantageous as quantum
coherence is very fragile and any incoherent interaction with a
macroscopic control apparatus can contribute to decoherence. Minimal
control is therefore likely to reduce controller-induced
decoherence, which is a limiting factor for many quantum information
tasks.  It has been shown that a wide class of coupled many-body
systems can be manipulated as required by controlling a local
subsystem~\cite{Lloyd}, and even for generic (not necessarily
multi-partite) systems, control of a small subspace or even single
transition is often sufficient to achieve full controllability
\cite{PRA78n062339}.  In particular, the XXZ-Heisenberg model is
controllable if one end spin can be fully
controlled~\cite{Burgarth-Bose}.

Controllability, however, only guarantees the existence of a control
to complete a certain task, but does not tell us how to generate
such a control, and even constructive controllability proofs do not
usually result in efficient solutions. Efficiency is critical due to
limited coherence time for many systems, which makes long pulse
sequences problematic.  Moreover, practical feasibility
considerations impose various constraints on the permissible fields,
from rise time to the complexity of its frequency spectrum.
Consequently, control design has to be considered case by case and
there is no general scheme that is optimal for an arbitrary system.
Recently, there have been several proposals for achieving certain
tasks in quantum information.  For instance, an efficient transfer
of quantum states can be achieved in an Ising chain, which is
accessible in Nuclear Magnetic Resonance (NMR), through global
\cite{fitzsimon} and mixture of global-local \cite{Khaneja} pulses.
In~\cite{Bur-Vit-Bose}, a realistic scheme is proposed to achieve
high fidelity state transfer by applying a sequence of two-qubit
gates at one end of the chain and it was shown that the two qubit
gates can be implemented by a proper sequence of switching the
interaction of the two spins.  Universal quantum computation along
the XY spin chain can be achieved by controlling the two spins at
one end of the chain~\cite{Burgarth,Kay}. In~\cite{Burgarth} one
spin is controlled by two independent magnetic fields and a fast
Hadamard gate operating on the other spin.  Alternatively, in
\cite{Kay} the two spins are controlled via three independent
control fields, two of which are non-local interactions.  Also, very
recently, almost perfect state transfer has been achieved by
controlling the magnetic field in the $z$ direction over all spins
in the ferromagnetic Heisenberg chain
\cite{giovannetti-fazio-montangaro}.

In this work, we consider the problem of entanglement generation
between the two ends of an anti-ferromagnetic Heisenberg spin chain
when control is restricted to a uni-directional local magnetic field
on one spin at either end of the chain. This type of control is not
sufficient for full controllability of the $N$-spin
system~\cite{Bur-Vit-Bose} as both the system and control
Hamiltonian are excitation preserving, and thus the system
decomposes in into $N+1$ excitation subspaces. For a Heisenberg
chain, we shall demonstrate that with the Heisenberg interaction and
the local control Hamiltonian, the system is controllable in the
particular subspace with the largest number of excitations, and thus
if the target state is in the same subspace as the initial state, it
is reachable by such local control. Through optimization, we find
the time-varying control field that generates an almost perfect
singlet pair between two ends of the chain, after the system is
initially prepared in the ground state. In general, calculations to
generate the optimal control field for the largest excitation
subspace of the anti-ferromagnetic chain are considerably more
demanding than the more common single-excitation subspace
calculations \cite{giovannetti-fazio-montangaro}, as the system
state lies in the $N/2$-excitation subspace for $N$ even, which has
dimension $N!/[(N/2)!]^2$ as opposed to dimension $N$ for the first
excitation subspace, and has a complicated form being a
superposition of many computational basis states. Considering this
scenario, however, has several benefits from a practical
perspective: (i) most realizations of spin chains are
anti-ferromagnetic in their nature~\cite{AFM}; (ii) in our
entanglement generation strategy the initial state is the ground
state of the system, which any system naturally takes when it cools
down and thus there is no need for extra control to initialize the
system; (iii) since the ground state of the system is highly
entangled it can be considered as an initial source for quantum
information tasks.  This helps simplify the control mechanism as
multipartite entanglement naturally exists and the control simply
has to convert it to utilizable bi-partite entanglement between the
distant end spins. Despite the highly restrictive nature of the
control and the dimension of the state space, we find that the
resulting controls are robust with regard to several non-ideal
features such as thermal fluctuations, decoherence, leakage of the
local magnetic field and existence of uncertainty in the couplings
of the system.

The paper is organized as follows: in Section II, we briefly introduce
the model and its invariant excitation subspace characterized by a
definite value of the total excitations. In Section III, we will
formulate the entanglement generation task into an optimization problem,
and numerically find the optimal control pulses. In Section IV, from
controllability point of view we provides some analysis trying to
understand why such proposed optimization process can generate the
required control. In Section V, we add some more practical conditions to
the problem and discuss how these extra conditions will affect the
control results.  The results are summarized in the section VI.

\section{System and Control Model}

We consider an isotropic Heisenberg chain of $N$ spin-$\frac{1}{2}$
particles with the nearest-neighbor interaction:
\begin{equation}
 \label{eqn:H0}
  H_s=JH_0=J\sum_{n=1}^{N-1} (X_n X_{n+1} + Y_n Y_{n+1} + Z_n Z_{n+1}),
\end{equation}
where the Pauli operators are defined as usual
\begin{equation*}
 X = \begin{bmatrix} 0 & 1 \\ 1 & 0 \end{bmatrix}, \quad
 Y = \begin{bmatrix} 0 & -i \\ i & 0 \end{bmatrix}, \quad
 Z = \begin{bmatrix} 1 & 0 \\ 0 & -1 \end{bmatrix}.
\end{equation*}
$J$ is the coupling between adjacent spins, and is positive for an
antiferromagnetic chain.  Let $\ket{0}$ and $\ket{1}$ be the basis
vectors of a single spin, representing the spin-down and spin-up
states. The $2^N$-dimensional Hilbert space $\H$ of the spin chain is
spanned by tensor products of the $N$ single spin basis vectors, the
computational basis states, for which we shall use the shorthand
notation $\ket{0\ldots 0} =\ket{0}^{\otimes N}$, etc.  For each basis
state the number of {\em excitations} is simply the number of $\ket{1}$s
in that basis.  For instance $\ket{11 \ldots 1}$ has $N$ excitations.
Setting $S_z=\sum_{n=1}^N Z_n$, it is easy to verify that $[H_0,S_z]=0$,
hence $H_0$ is invariant on every excitation subspaces.  Accordingly,
$\H$ can be written as a direct sum of $N+1$ excitation subspaces $\H_n$
($n=0,\ldots,N$) whose basis vectors are those having $n$
excitations. The dimension of $\H_n$ is $\binom{N}{n}$.  For chains with
an even number of spins the system~(\ref{eqn:H0}) has a unique ground
state in the $\frac{N}{2}$-excitation subspace $\H_{N/2}$.  For
odd-length chains the ground state is doubly degenerate with one ground
state in the $\H_{(N-1)/2}$ and $\H_{(N+1)/2}$ subspace each.

Next, we impose a local magnetic field along the $z$-direction on the
left end spin, with the resulting Hamiltonian:
\begin{equation}\label{eqn:H1}
H_c=B(t)H_1=B(t)Z_{1}
\end{equation}
where $B(t)$ is the magnitude of $H_c$, which can be varied over time.
For spin chains comprised of coupled quantum dots, such an interaction
could be achieved using local voltage gates. This is the Hamiltonian we
shall control for entanglement generation. The direction of field is not
essential due to the symmetry of the Heisenberg Hamiltonian. The main
idea is to demonstrate that perfect end-to-end entanglement can be
generated from only a local magnetic field along a given direction.
This requirement is experimentally simpler than other proposals where
independent control fields along different directions are required.  Of
course, the more controllable degrees of freedom, the more types control
tasks can be completed, but a single unidirectional local field is
sufficient for entanglement generation, and indeed controllability on
the $N/2$-excitation subspace as we shall see.

The total Hamiltonian of the spin chain thus becomes
\begin{equation}
 \label{eqn:H}
  H(t)=H_s+H_c=JH_0+B(t)H_1.
\end{equation}
Since $H_1$ is invariant on every excitation subspace $\H_M$, $H(t)$ is
also invariant on these subspaces.  As a result the number of
excitations of a given initial state $\ket{\psi(t)}$ does not change
under the evolution
\begin{equation}
 \label{eqn:state}
 \frac{d }{dt} \ket{\psi(t)} = -iH(t) \ket{\psi(t)},
\end{equation}
i.e., if $\ket{\psi(0)}$ is in the $M$-excitation subspace then
$\ket{\psi(t)}$ will be in $\H_M$ for all $t$.  Throughout this paper we
choose units such that $\hbar=1$, so $J$ and $B(t)$ has the dimension of
frequency.  For NMR experiments $J$ is typically a few hundred Hz, and
$B(t)$ is around 50 kHz for liquid-state NMR and a few hundred kHz for
solid-state NMR~\cite{Chuang}.

\section{Optimal Control Design}

Our question is whether we can find an appropriate control pulse $B(t)$
in the model~(\ref{eqn:H}) to generate a maximally entangled state
between the two end spins at a certain final time $t=t_f$.  We shall
first assume that the system is initialized in the ground state
$\ket{G_0}$ of $H_0$ and thus in a pure state.  In principle, this can
be done by simply cooling the systems and this assumption will later be
relaxed.  As mentioned above, for even chains the ground state of $H_0$
is unique and lies in the subspace of $\H_{N/2}$ so that cooling the
system is enough to prepare it.  For chains of odd length the degeneracy
can be lifted by applying a small global magnetic field during the
cooling process and the system can be prepared into the ground state in
the subspace $\H_{(N+1)/2}$. In the following, for simplicity, we will
call $\H_{N/2}$ for even chain and $\H_{(N+1)/2}$ for odd chain as the
largest excitation subspace, and we assume the system is initially
prepared as the ground state $\ket{G_0}$ in the largest excitation
subspace.

The control objective is to find a time-varying control field $B(t)$,
$t\in[0,t_f]$, such that the final state at time $t_f$ takes the form
\begin{equation}
 \label{eqn:psit}
 \ket{\psi(t_f)} = \ket{\psi^-}_{1,N}\otimes \ket{\psi}_{\rm re},
\end{equation}
where $\ket{\psi^-}=(\ket{01}-\ket{10})/\sqrt{2}$ is the singlet state
between the first and the last spin, and $\ket{\psi}_{\rm re}$ is the state
of the remainder, i.e., spins $n=2,3,...,N-1$, on which we do not impose
any constraint.

One of the methods to generate the control pulse $B(t)$ is through
optimization, where the target state $\ket{\psi(t_f)}$ is the maximal
(or minimal) point of a certain objective function $K[B(t)]$.  A good
candidate in our case is
\begin{equation}
 \label{eqn:func}
  K[B(t)]=\bra{\psi(t_f)} A \ket{\psi(t_f)},
\end{equation}
where the operator $A$ is the projection onto the target subspace
\begin{equation}
 \label{eqn:A}
  A=|\psi^-\ra_{1,N}\la \psi^-| \otimes I_{\rm re},
\end{equation}
and $\ket{psi^-}_{1,N}\bra{\psi^-}$ is the density operator for the
singlet state between spins $1$ and $N$, which we desired to prepare,
and $I_{\rm re}$ is the identity operator acting on the rest part of the
chain.  Eq.~(\ref{eqn:func}) shows that this function is the
\emph{fidelity} of the density matrix of the end spins $\rho_{1,N}(t)$
with respect to the singlet state at time $t_f$. Consequently, this
function is always positive and assumes its maximum for states of the
form (\ref{eqn:psit}), where the two end spins form a singlet.  Hence,
our main goal is to drive this system to its maximum by varying $B(t)$
over the time interval $[0,t_f]$.  For simplicity, we will vary $B(t)$
in the function space of the piecewise-constant functions.

The optimization procedure is as follows.  We divide the time $[0,t_f]$
into $p$ equally spaced sub-intervals, $\Delta {t}=t_f/p$.  On each time
interval the control field $B(t)$ takes the constant value $B_m$
($m=1,2,..,p$) satisfying $|B_m|<B_M$, where $B_M$ is the maximum
amplitude of the control field that we can generate experimentally.
Assuming unitary evolution governed by the Schr\"odinger
equation~(\ref{eqn:H}) the final state subject to this piece-wise
constant control is simply
\begin{equation}
\label{eqn:evolution}
  \ket{\psi(t_f)} = U(B_p)\cdots U(B_1) \ket{G_0},
\end{equation}
where
\begin{equation}
\label{eqn:U}
  U(B_m) = e^{-i(J H_0+B_m H_1)\Delta t},
\end{equation}
and the fidelity is simply a function of $p$ variables, $K[B(t)]=
K(B_1,\dots,B_p)$.  This is a standard optimization problem in a
finite-dimensional parameter space, which can be solving using various
methods.  We use a quasi-Newton method developed by Boyden, Fletcher,
Goldfarb and Shanno~\cite{BFGS} (BFGS), which is an improvement over the
standard GRAPE algorithm~\cite{GRAPE} widely used for quantum control
problems.

The algorithm is summarized as follows: (1) Choose an initial trial
function $B(t)=B^{(1)}(t)$, which could be constant $B^{(1)}(t)=1$ or
some other random choice. (2) Iteratively generate $B^{(m+1)}(t)$ from
$B^{(m)}(t)$ by applying the quasi-Newton method, which guarantees
$K[B^{m+1}(t)]>K[B^{m}(t)]$.  The algorithm is terminated if the change
of the fidelity is smaller than a given threshold, or we cannot find a
search direction to increase the fidelity further.  The method
guarantees that the function sequence $B^{(m)}(t)$ converges to a local
maximum $B_{max}(t)$ of the function $K[B(t)]$, and constraints, e.g.,
on the field amplitudes, can be explicitly taken into account.  In
applications such as NMR where the field amplitudes $B(t)$ can be large
compared to $J$ (at least $10^2-10^3$~J) it is usually not necessary to
impose explicit amplitude constraints --- at least our simulation
results for this problem suggest that the amplitudes of the optimized
fields are usually well below $100$~J if we start with an initial field
$B^{(1)}(t)$ close to the zero field $B\equiv 0$, as is to be expected
for a local optimization method.  Therefore in most of the following
calculations no amplitude constraints have been imposed, except where
otherwise stated.  In general amplitude constraints increase the amount
of time required to reach the objective but the imposition of such
constraints can be beneficial by increasing the robustness of the
optimal pulses obtained with regard to noise, field leakage, etc.
Beside the fidelity $K[B(t)]$ of the final state with respect to the
desired singlet state, we also use the entanglement measured by the
concurrence~\cite{wootters} to quantify the quality of our procedure,
where the entanglement $C$ is maximal ($C=1$) if we successfully
generate a perfect singlet.

\begin{figure}
\includegraphics[width=\columnwidth]{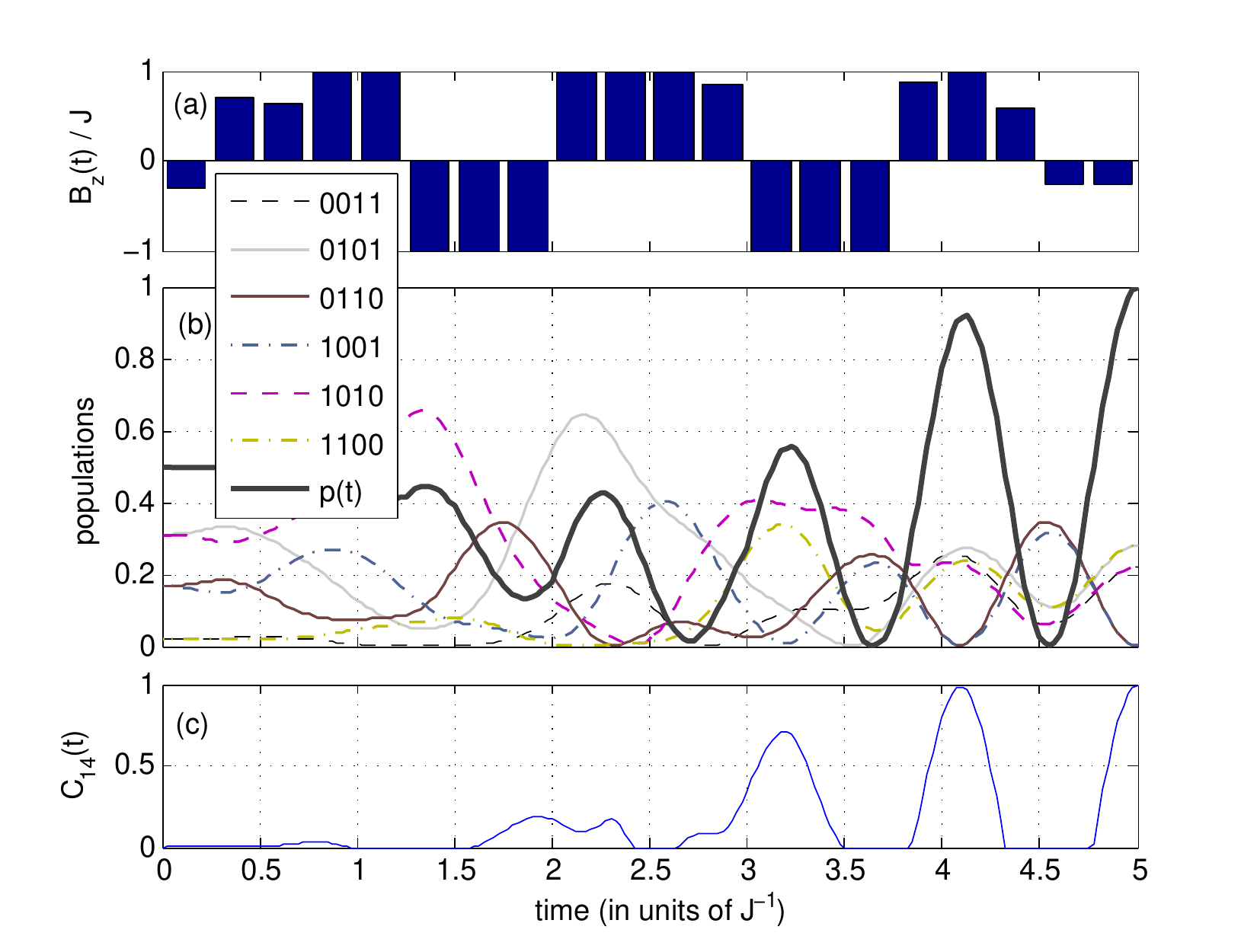} \caption{(Color online)
(a) Example of optimal field $B_z(t)$ for $N=4$ and (b) corresponding
evolution of the populations of the computational basis states in the
$2$-excitation subspace and the population $p(t)$ of the state
$\ket{\psi} =\frac{1}{2}[\ket{0011}-\ket{0101}-\ket{1010}+\ket{1100}]$
(thick solid line) and (c) concurrence $C_{14}(t)$ between spins $1$ and
$4$.  The evolution shows that the pulse transfers virtually all
population to the state $\ket{\psi}$, which exhibits perfect
entanglement between the end spins, as demonstrated by the concurrence
of the reduced density matrix.  (Pulse shown obtained with amplitude
constraint $|B(t)|/J\le 1$.)} \label{fig:evolution}
\end{figure}

Using this optimal control framework we computed optimal pulses for spin
chains of length $N=4$ to $20$.  An example of an optimal control field
and the corresponding evolution of the system for $N=4$ is shown in
Fig.~\ref{fig:evolution}.  The control steers the ground state of $H_0$,
which is a superposition of all six computational basis states in the
$N/2=2$ excitation subspace $\H_2$, to a $+1$-eigenstate of the
observable $A$ in the same subspace.  For $N=4$ the $+1$-eigenspace of
$A$ restricted to $\H_2$ is spanned by
$\{\frac{1}{\sqrt{2}}(\ket{0011}-\ket{1010}),\frac{1}{\sqrt{2}}
(\ket{0101}-\ket{1100})\}$ and the control must steer the initial state
to some state in this subspace.  In the example shown in
Fig.~\ref{fig:evolution}, for instance, virtually all the population is
transferred to the eigenstate
$\ket{\psi}=\frac{1}{2}(\ket{0011}-\ket{1010}-\ket{0101}+\ket{1100})\}$
at the final time, although the population of the target state and the
concurrence between the end spins do not increase monotonically unlike
for some local optimization approaches~\cite{PRA80n042305}.  The partial
trace of the final state over the inner spins (up to three digits
accuracy) is
\begin{equation*}
  \rho_{14}= \Tr_{2,3}(\ket{\psi}\bra{\psi})
           =  \begin{pmatrix}
          0 &  0 &  0 & 0 \\
          0 &  0.501 & -0.500 & 0 \\
          0 & -0.500 &  0.499 & 0 \\
          0 & 0 & 0 &
         \end{pmatrix},
\end{equation*}
which is an almost perfect Bell state with concurrence $0.99998$, and we
can achieve fidelities and concurrences almost arbitrarily close to $1$
by relaxing the constraints on the field amplitudes, pulse duration or
time resolution of the fields slightly.  In this sense we claim that
this method can generate perfect entanglement.

\begin{figure}
\includegraphics[width=\columnwidth]{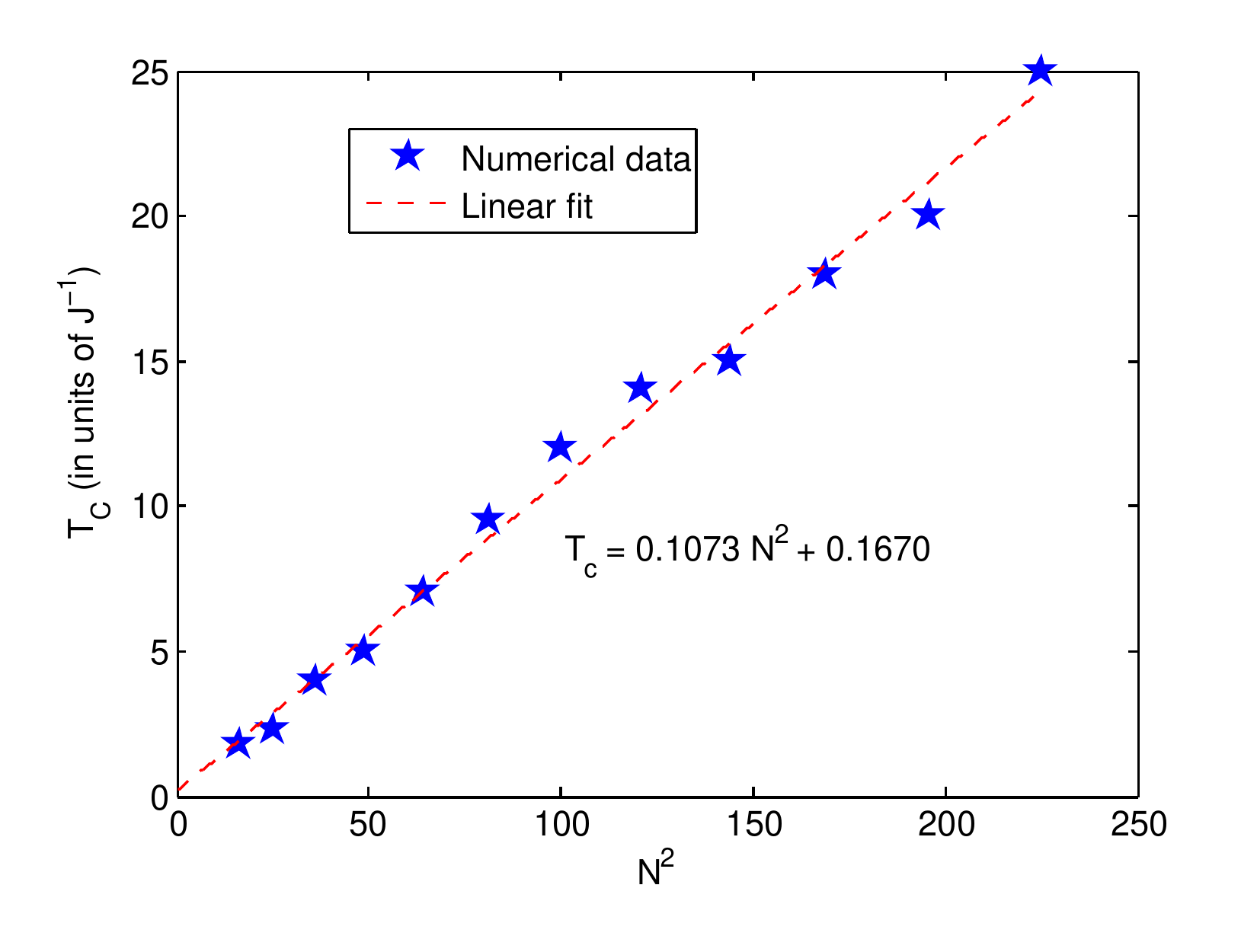} \caption{(Color online)
Lower bound on the time $T_C$ required for generating entanglement in
terms of $N^2$, based on numerical optimization results for chains of
lengths $N=4$ to $15$ without constraints. } \label{Fig_Time}
\end{figure}

Although the dimension of the $\frac{N}{2}$-excitation subspace
increases from $\binom{4}{2}=6$ for $N=4$ to $\binom{20}{10}=184,756$
for $N=20$, rendering the optimization problem considerably more
challenging and time-consuming, we are still able to find optimal
optimal pulses for which $K[B]$ assumes values very close to $1$ in
relatively short time, suggesting that our optimization method for
generating end-to-end entanglement is still effective for relatively
long spin chains.  For a given chain there is always a lower bound $T_C$
for the evolution time $t_f$, however, below which we can never achieve
unit fidelity no matter how many time steps or iterations are
used. Unfortunately, analytical expressions for these bounds are known
only for very simple systems. We therefore numerically explore the
search space to find approximate lower bounds $T_C$ as a function of the
chain length $N$.  The results, shown in Fig.~\ref{Fig_Time} suggest a
quadratic increase with respect to chain length $N$, similar to what was
reported in~\cite{Burgarth} for implementing the universal quantum
computation by controlling two qubits of an XY chain, rather than a
linear increase as in some other entanglement generating schemes
\cite{bayat-xxz,bayat-quench} where entanglement propagates from one end
through the chain.  The nonlinear dependence on the length of the chain
is not surprising as the mechanism for increasing the entanglement
between the end spins in our scheme involves dynamic population transfer
from the ground state of $H_0$ to an eigenstate of the observable
$A$. We note that the control itself cannot create entanglement directly
as the control Hamiltonian $H_1$ is completely local.  It can only alter
the entanglement between the end spins by taking advantage of the
existing interactions between all spins and the resulting non-trivial
dynamics.

\section{Controllability \& Reachability Considerations}

Considering the restrictions imposed on the control, it is rather
surprising that we are able to find optimal control solutions at
all, and it is worthwhile to investigate whether there always exist
controls that achieve perfect entanglement, at least in principle.
In order to answer these questions, we need to review the concepts
of reachability and controllability.  Ultimately, the question we
would like to answer is what is the set of states that is reachable
from the ground state for a system governed by the evolution
(\ref{eqn:state}). We already know that there are some constraints.
For a Hamiltonian system the spectrum of the density operator is
invariant.  For pure initial states this simply means that the set
of reachable states is a subset of the pure states. We also know
that there are additional limitations as both $H_0$ and $H_1$ in our
case commute with total excitation operator $S_z$, and hence the
Hilbert space decomposes into excitation subspaces on which the
dynamics is invariant.  Thus if the initial state is a pure state on
the $\frac{N}{2}$ excitation subspace then the set of reachable
states is a subset of the pure states on this excitation subspace.
However, the key question is whether the target state
(\ref{eqn:psit}) in this excitation subspace is reachable from the
initial state $\ket{G_0}$. Unfortunately, this is a very difficult
question to answer directly in general, except for controllable
systems.  We must be careful here as there are different degrees of
controllability~\cite{JPA35p4125} but if we can show, for instance,
that the unitary evolution of the system is controllable, then
pure-state controllability follows.

The Schr\"odinger equation (\ref{eqn:state}) can be rewritten into
an equation in terms of the unitary process $U(t)$ as the variable
\begin{equation}
\label{unitary} 
 \dot U(t) = -i(J H_0+B(t)H_1)U(t)
\end{equation}
where $U(0)=I$ and the solution for a given control $B(t)$ is denoted as
$U(t,B(t))$.  In this context, the reachable set $\RR$ is defined as the
set of unitary matrices that can be achieved by the dynamics $U(t,B)$
for some (admissible) control $B(t)$, i.e., $\bar U\in \RR$ if and only
if there exists an admissible control $B(t)$ such that $U(t_f,B(t))=\bar
U$ and we have the following~\cite{Jurdjevic}:

\begin{theorem}
Let $\L$ be the Lie algebra generated by $\Span\{-iH_0,-iH_1\}$, often
called the dynamical Lie algebra.  Then $\RR=e^{\L}$.
\end{theorem}

Thus, the reachable set depends on the dynamical Lie algebra generated
by the Hamiltonian.  If $\L=\uu(2^N)$ or $\L=\su(2^N)$, then
$\RR=\UU(2^N)$ or $\RR=\SU(2^N)$ and the system is \emph{controllable}
in the sense that we can implement any unitary process up to a global
phase and steer any density operator to any other density operator with
the same spectrum.  Back to our spin chain problem, for $H=J
H_0+B(t)H_1$ with $H_0$ and $H_1$ as in Eqs (\ref{eqn:H0}) and
(\ref{eqn:H1}), if the system is controllable on $\H_{N/2}$ for $N$ even
or $\H_{(N\pm 1)/2}$ for $N$ odd, then the existence of the control to
generate the required singlet state is guaranteed.  A straightforward
way to check controllability is by calculating the dimension of the
dynamical Lie algebra.  Calculations for small chains of different
length suggest full controllability on the relevant excitation subspaces
for chains of both even and odd length.  However, such calculations are
laborious and impractical for long chains.  Similarly, proving that the
dimension of the dynamical Lie algebra is $M^2$, where $M$ is the
dimension of the $\H_{N/2}$ or $H_{(N-1)/2}$ excitation subspace, for
any $N$ is a difficult mathematical problem.  However, for even-length
isotropic Heisenberg chains, controllability can be easily verified
using the following useful result~\cite{Altafini}:

\begin{theorem}
\label{regular} If $H_0$ is effectively strongly regular and $H_1$ is
connected, then the system is controllable.
\end{theorem}

Here $H_0$ is strongly regular if all the transition frequencies
$\omega_{k\ell}=e_k-e_\ell$, where $e_k$ are the eigenvalues of $H_0$,
are distinct.  Choosing a basis such that $H_0$ is diagonal,
$H_0=\diag(e_1,e_2,\ldots,e_n)$, let $H_1=(b_{ij})$ be the matrix
representation with respect to this basis.  $H_1$ is connected if the
graph represented by $H_1$ is a connected graph.  To be more specific,
we say the indices $j$ and $k$ are connected if $b_{jk}\ne0$, and if
every two indices are connected through several nonzero $b_{jk}$'s, then
$H_1$ is called connected.  Moreover, $H_0$ is called effectively
strongly regular, if $\omega_{jk}$, $j\ne k$, are nonzero and distinct
whenever $b_{jk}\ne 0$.  Using these definitions and the previous
theorem we can verify on a case by case basis that even-length
Heisenberg chains are controllable on $\H_{N/2}$.

\begin{example}
For $N=4$ we have $\dim(\H_2)=6$, $H_0$ and $H_1$ take the following
forms in the eigenbasis of $H_0$:
\begin{align*}
H_0&=\begin{pmatrix}
-2\sqrt{3}-3  &  0 &   0 &  0 &  0 &   0\\
    0  &  -2\sqrt{2}-1 &   0 &  0 &  0 &   0\\
    0  &  0 &   -1 &  0 &  0 &   0\\
   0  &  0 &   0 &  2\sqrt{3}-3 &  0 &   0\\
   0  &  0 &   0 &  0 &  2\sqrt{2}-1 &   0\\
   0  &  0 &   0 &  0 &  0 &   3\\
\end{pmatrix},\\
H_1 &=\begin{pmatrix}
    0    &  0.81 &  0.58 &  0    &  0.11 &  0   \\
    0.81 &  0    &  0    &  0.50 &  0    & -0.31\\
    0.58 &  0    &  0    & -0.58 &  0    &  0.58\\
    0    &  0.50 & -0.58 &  0    & -0.65 &  0   \\
    0.11 &  0    &  0    & -0.65 &  0    & -0.75\\
    0    & -0.31 &  0.58 &  0    & -0.75 &  0
\end{pmatrix},
\end{align*}
which shows that $H_0$ is effectively strongly regular and $H_1$ is
connected.  Therefore, the system is controllable on the $\H_2$
excitation subspace, and by virtue of controllability any pure state
in the $\H_2$ excitation subspace is reachable from the ground state
$\ket{G_0}$, including our target Bell state (\ref{eqn:psit}).
\end{example}

We can similarly verify that the systems governed by (\ref{eqn:H})
satisfy the criteria for controllability above on $\H_{N/2}$ for $N$
even ranging from $4$ to $16$. For longer chains, such verification
will become inefficient again due to the evaluations of eigenvalues
and eigenvectors for large matrices.  For odd $N$, $H_0$ ceases to
be effectively strongly regular on the $(N\pm 1)/2$ excitation
subspace for $N>3$, and the above theorem does not apply.  However,
this does \emph{not} imply non-controllability as the above
controllability criteria is only sufficient but not necessary.  In
fact direct calculation of the dimension of the dynamical Lie
algebra for odd chains suggests that the system is still
controllable.  For instance, for $N=5$ we obtain $\dim \H_{2}
=\binom{5}{2}=10$ and direct computation of the Lie algebra shows
that it has dimension $100=10^2$, suggesting that the Lie algebra in
this case is $\uu(10)$, which implies full controllability on the
subspace.

Although controllability implies reachability and the existence of a
control that steers the initial state to the target state, it should
be noted that it does not answer the question of the minimal time
required to steer the system to the target state. Furthermore, in
practice the finite time resolution of the field and bounds on the
field strengths impose additional constraints, i.e., even if there
exists an unbounded control that steers the initial state to the
target state in time $t_f$, there may not be a piecewise constant
solution of the form $(B_m)=(B_1,\ldots,B_p)$ if $p$ is fixed and
$|B_m|\le B_M$ for some fixed bound $B_M$. Finally, most
optimization techniques including the quasi-Newton-method used in
the previous section are only guaranteed to find some extremum of
the objective function, which need not necessarily correspond to a
global optimum.

For more general spin chain models, for example, the XXZ chain, we
can still use the above optimization method to generate perfect
entanglement.  For different length of chains we have calculated,
the system is always found to be controllable on the subspace in
question. More interestingly, for XX chain, such optimization method
can also generate perfect entanglement from the ground state, but
the system is found to be non-controllable on that subspace. This
gives an example where the perfect end-to-end entangled state is
reachable from the ground state, but controllability does not hold,
a strong evidence that reachability is weaker than controllability.

\section{Control Performance under Realistic Conditions}

In the previous sections we have illustrated that a single magnetic
field in the $z$-direction acting on a single spin suffices to
generate a perfect entanglement between two ends of a Heisenberg
chain. However, the optimal control design is based on four ideal
assumptions, namely that
\begin{enumerate}
\item The system can be prepared in its ground state;
\item The control field can locally address a single spin without
      affecting the others;
\item The system is homogeneous, i.e., all couplings are equal and
      the coupling strength is known;
\item The system is isolated from its environment.
\end{enumerate}
In this section we consider how the effectiveness of the optimal
controls is affected when these assumptions are relaxed.

\subsection{Thermal Initial State}

\begin{figure}
\includegraphics[width=\columnwidth]{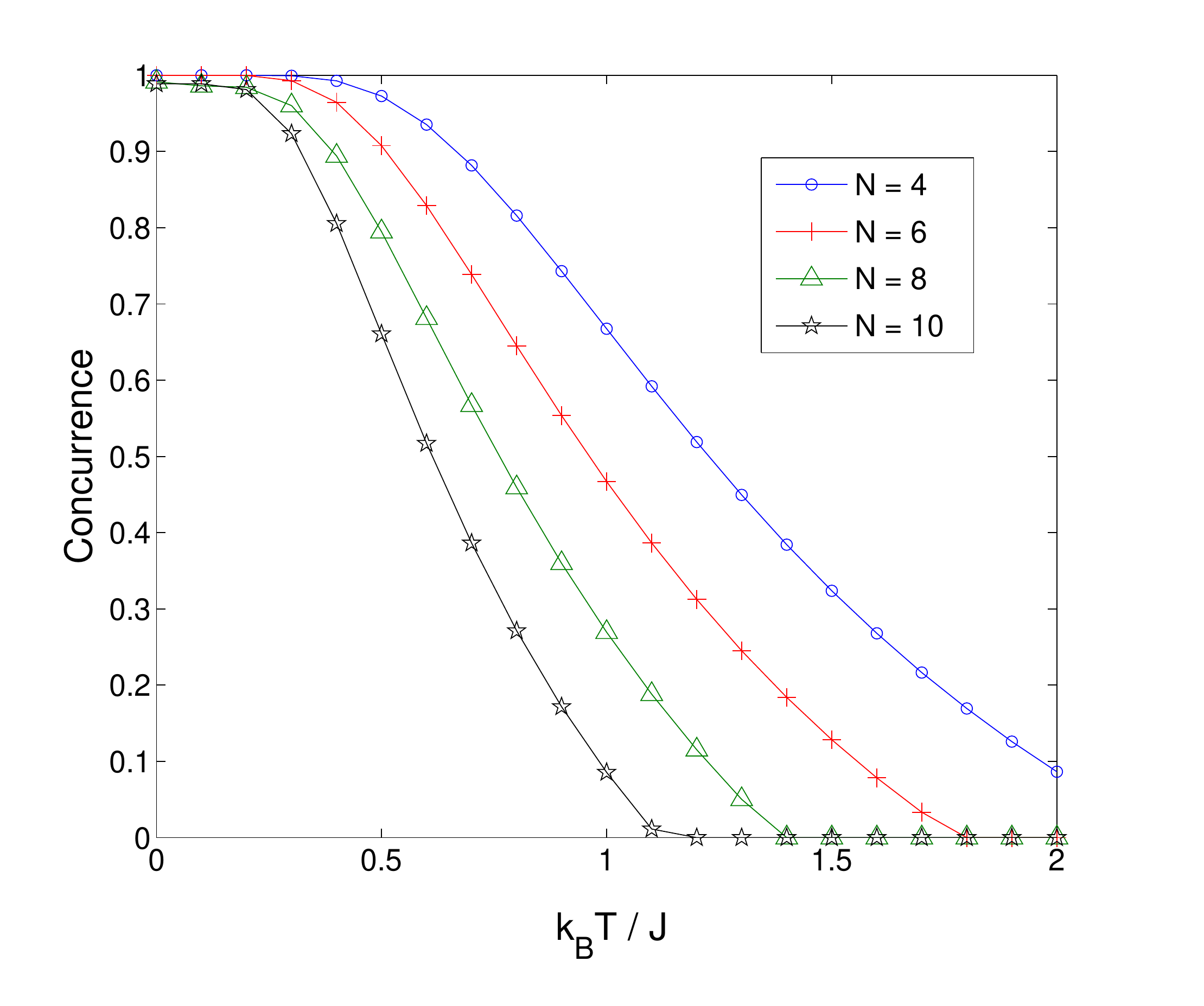} \caption{(Color online)
Final concurrence between end spins vs $k_BT/J$ for chains of different
lengths if the pulse optimized for entanglement generation from the
ground state is used.  For large $k_BT$ the concurrence approaches zero
and the pulse is ineffective, but for $k_BT/J\ll 1$ (low temperature
limit) the concurrence remains close to unity.}  \label{Fig_Temp}
\end{figure}

In an ideal setup we would be able to cool the system to its ground
state $\ket{G_0}$ but in practice this is difficult, and thus the
initial state is likely to be a thermal ensemble
\begin{equation}
  \label{rho0}
  \rho(0) = \frac{e^{-H_0/k_BT}}{\Tr[e^{-H_0/k_BT}]},
\end{equation}
where $T$ is temperature and $k_B$ is the Boltzmann constant, which
contains some contribution of the excited states, including states
outside the subspace $\H_{N/2}$.  To assess the effectiveness of the
optimal control pulse generated for the ground state in this case,
we evolve the initial state (\ref{rho0}) under the action of the
total Hamiltonian $H(t)$ given in Eq. (\ref{eqn:H}) subject to the
pulse $B_{max}(t)$ optimized for the ground state,
\begin{equation}
  \label{rhot}
  \rho(t_f)= \exp_+\left[-i\int_0^{t_f} \!\!\! H(t) dt \right]
             \rho(0)
             \exp_+\left[-i\int_0^{t_f}\!\!\! H(t)\, dt \right]^\dag,
\end{equation}
where $\exp_+$ indicates positive time ordering, and compute the
density matrix for the end spins $\rho_{1N}(t_f)$ by tracing out the
interior spins.  Fig.~\ref{Fig_Temp} shows the resulting
entanglement as measured by the concurrence as a function of $k_B
T$.  For sufficiently low temperatures $T$ the optimal control field
still generates good entanglement. For example, for a chain of
length $N=10$ entanglement survives up to $k_BT/J\simeq 1.2$.
Moreover, Fig.~\ref{Fig_Temp} shows that for low temperatures the
concurrence is rather flat, implying that up to a certain
temperature the control performance is minimally affected by
thermalization.  The width of the plateau is related to the energy
gap between the ground state and the first excited state.  For
thermal energies $k_B T$ less than this energy gap the population of
the excited states remains very small, and only when $k_BT$ exceeds
this limit other eigenstates become populated.  Since the energy gap
between the ground state and first excited states decreases with
chain length, the width of the plateau also decreases.

\begin{figure}
\includegraphics[width=\columnwidth]{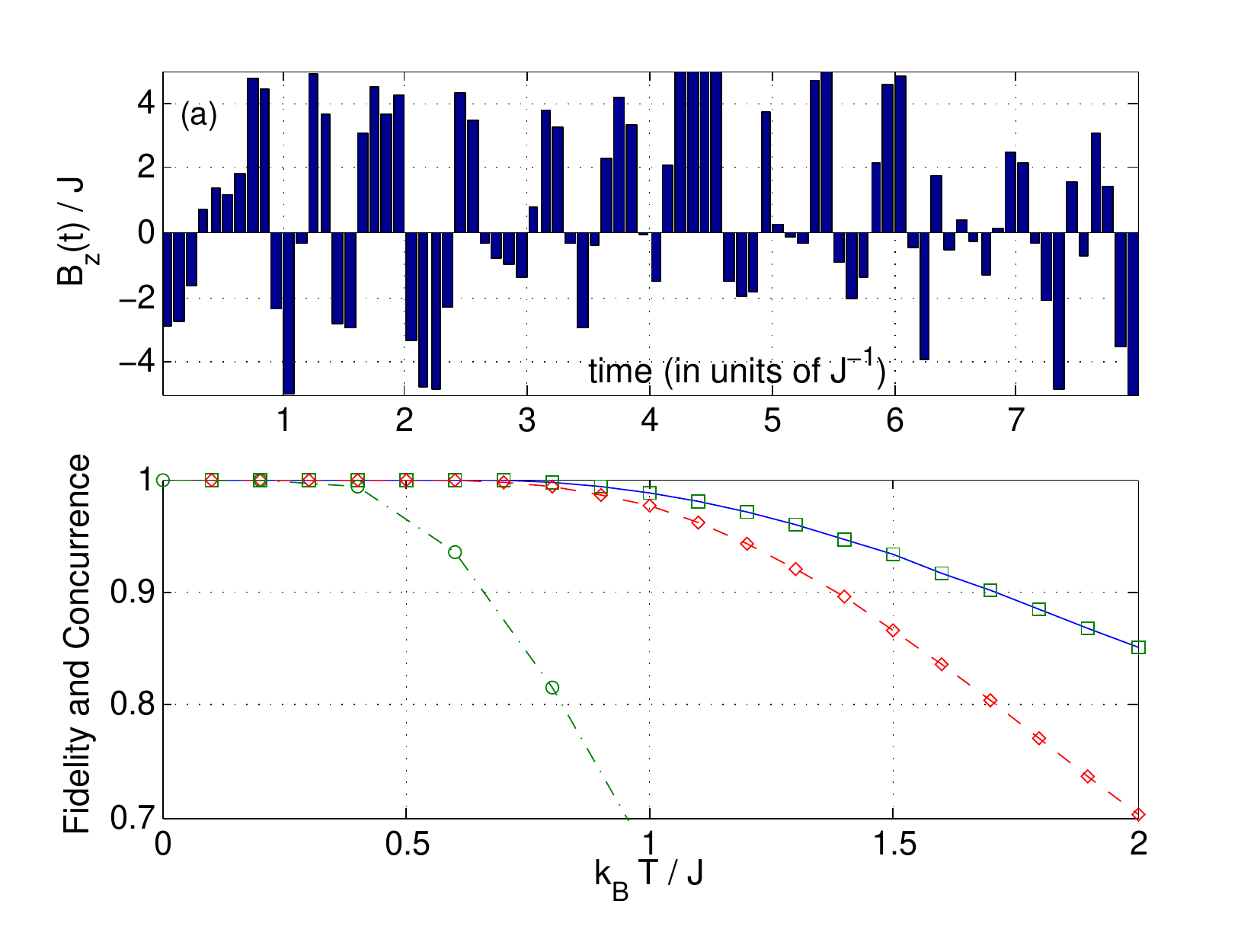}
\caption{(Color online) (a) Optimal control pulse $B_z(t)$ obtained by
optimizing $\Tr[A\rho(t_f)]$ for an initial thermal ensemble with
$k_BT/J=2$ for $N=4$. (b) If if this pulse is applied to initial
ensembles $\rho(0)$ with $k_BT/J$ ranging from $0$ to $2$, the fidelity
$\Tr[A\rho(t_f)]$ achieved (squares) assumes its theoretical upper bound
(solid blue line), showing that this field is optimal in terms of
maximizing $\Tr[A\rho(t_f)]$ for initial ensembles over a wide range of
temperatures.  Accordingly, the corresponding entanglement as measured
by the concurrence of the reduced density matrix $\rho_{14}(t_f)$ for
the ensemble-optimized pulse (dashed red line with diamonds) is
substantially higher than that for a ground state optimized pulse
(green dash-dot line with circles).}  \label{Fig_Temp2}
\end{figure}

These results are consistent with expectations.  For low temperatures
the ground state of the system dominates the mixture of the thermal
state (\ref{rho0}), and the optimal control pulse transfers this to a
$+1$-eigenstate of $A$ that has unit concurrence.  Most of the other
states of the initial ensemble will be mapped to $0$-eigenstates of $A$
and not contribute to the entanglement between the end spins, but if the
initial populations of the excited states are small, the pulse will
remain mostly effective. The results can be improved greatly if we
calculate the optimal pulse for the initial thermal state, rather than
using the optimal control for the ground state, i.e., maximizing the
fidelity $\Tr[A\rho(t_f)]$ starting with a thermal ensemble $\rho(0)$,
although there are upper bounds on the maximum achievable fidelity in
this case.  Let $\rho(0)=\oplus_{n=0}^N \rho^{(n)}(0)$ where
$\rho^{(n)}(0)$ is the initial state restricted to the $n$th excitation
subspace, $w_{n,m}$ be the populations of the initial ensemble on the
$n$th excitation subspace with $w_{n,m}\ge w_{n,m+1}\ge 0$ for
$m=1,\ldots,\dim\H_{n}$, and let $d_n$ be the number of $+1$-eigenvalues
of $A$ on the $n$th excitation subspace.  Then the constraint of unitary
evolution combined with the fact that the dynamics is invariant on each
excitation subspace imposes the following upper bound on the maximum
achievable fidelity~\cite{Bounds}
\begin{equation}
   \label{eqn:KB}
   \Tr[A \rho(t_f)] \le \sum_{n=0}^N \sum_{m=1}^{d_n} w_{n,m}.
\end{equation}
Fig.~\ref{Fig_Temp2} shows that we can effectively realize this upper
bound for a wide range of $k_BT$.  In fact, the figure shows that it is
not necessary to know the initial temperature.  The pulse that achieves
the maximum fidelity possible for $k_B T/J=2$ is an optimal solution for
\emph{all} temperatures in that it achieves the maximum possible value
of $\Tr[A\rho(t_f)]$ for every initial thermal ensemble $\rho(0)$.  This
can be explained as follows.  Denoting the eigenstates of $H_0$ by
$\ket{e_{n,m}}$ for $n=0,\ldots,N$ and $m=1,\ldots,\dim\H_n$, the
initial state can be written as
\begin{equation}
   \sum_{n=0}^N \sum_{m=1}^{\dim\H_n} w_{n,m} \ket{e_{n,m}}\bra{e_{n,m}},
\end{equation}
where we have ordered the spectrum and the corresponding eigenstates
in an non-increasing way for both $\rho(t_f)$ and $A$.  Different
thermal ensembles at different temperature have different spectra
$w_{n,m}$. However, to realize the upper bound~(\ref{eqn:KB}), any
optimal pulse for a given thermal ensemble must map all the
eigenstates $\ket{e_{n,m}}$ to the eigenstates of $A$ in the same
order, independent of the actual value of $w_{n,m}$.  Therefore, the
optimal pulse for a thermal ensemble with $T_1>0$ is also an
optimal pulse for any thermal ensemble with $T_2<T_1$, and in
particular, it is an optimal pulse for the initial ground state.
But the inverse is not true: the optimal pulse obtained if we start
in the ground state is in general not optimal for an initial thermal
ensemble.  This is because the optimal field starting in the ground
state is only guaranteed to map the final state onto one of the 
$+1$-eigenstates of $A$, but it generally will not map the correct 
excited states onto other $+1$-eigenstates of $A$, and therefore 
it is not an optimal pulse for an initial thermal ensemble. This is 
illustrated in
Fig.~\ref{Fig_Temp2}, which shows that the fidelity that can be
achieved if a ground-state optimized pulse is applied to a thermal
ensemble with $T>0$ is strictly smaller than the optimal fidelity
that is achievable with a pulse optimized for this thermal ensemble.
The concurrence exhibits a similar behavior.  For $N=4$ the final
concurrence between the end spins for $k_BT/J=1$ is $0.9762$ vs less
than $0.7$ for the ground-state optimized pulse in
Fig.~\ref{Fig_Temp}.

\subsection{Local control with Gaussian leakage}

Another central assumption of our control scheme is that the control
field is confined and interacts only with the first spin.  In
practice such perfect confinement is difficult to achieve and any
local magnetic field will also affects neighboring spins, although
the effect will usually decay with increasing distance from the
control spin, the first spin in our case, either exponentially or
following a power law.

\begin{figure}
\includegraphics[width=\columnwidth]{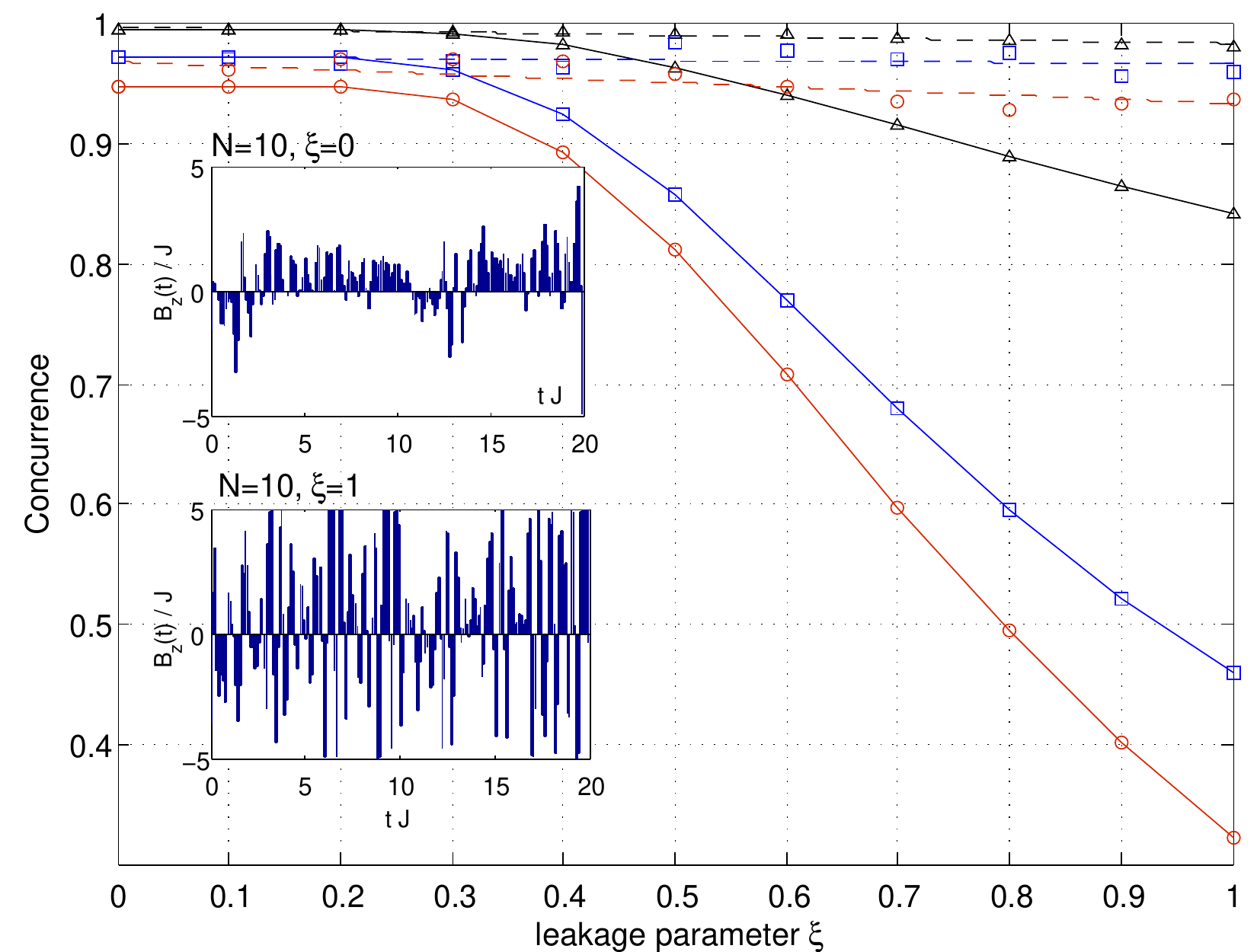}
\caption{(Color online) Entanglement as a function of leakage $\xi$ for
different chain length ($N=6$ black triangles, $N=8$ blue squares,
$N=10$ red circles).  Solid lines indicate the concurrence dependence
for pulses optimized assuming ideal $H_1$ and dashed lines indicate
concurrence behavior for pulses optimized with leakage taken into
account, i.e., $\bar{H}_1$.  For $\xi$ small the optimal pulses obtained
assuming no leakage still achieve high concurrence but if $\xi$ can be
estimated, the entanglement generated can be substantially improved by
taking leakage into account in the optimization.  The inset shows that
the optimal pulses in the presence of leakage can differ substantially
from the optimal pulses in the absence of leakage. Note that optimal
control solutions are not unique and the robustness of individual
solutions with regard to leakage varies.}  \label{Fig_leak}
\end{figure}

To assess the effect of leakage on the performance of the control scheme
we take the pulses optimized under ideal assumptions and consider what
happens if the field decays exponentially with distance thus affecting
not only the target spin but also neighboring spins, effectively
changing the control Hamiltonian to
\begin{equation}
 \label{H_leak}
  H_c = B(t)\bar{H}_1=B(t)\sum_{k=1}^N e^{-(k-1)/\xi}Z_k
\end{equation}
where $\xi$ is a parameter quantifying the field leakage with
$\xi=0$ corresponding to the ideal case where the field is
completely localized and affects only the first spin.  Leakage
increases with $\xi$ and for $\xi \rightarrow \infty$ the control
field is uniform, resulting merely in a dynamic global energy shift
or global phase.  Fig.~\ref{Fig_leak} shows the effect of leakage on
the entanglement between the end spins at the final time as a
function of the parameter $\xi$.  As expected, entanglement decays
with increasing $\xi$ but even for $10\%$ leakage we still have a
good entanglement.  Furthermore, if the leakage parameter $\xi$ is
approximately known, we can again incorporate this information into
the optimization, replacing $H_1$ by $\bar{H}_1$ and then doing the
optimization. This substantially improves the results, allowing us
to achieve consistently high entanglement for a wide range of $\xi$,
although for very large $\xi$ $\bar{H}_1$ almost commutes with
$H_0$, rendering the control increasingly ineffective (see Fig.
\ref{Fig_leak}).

\subsection{Random Couplings}

Another assumption in our model-based optimal control scheme is that the
system Hamiltonian $H_0$ is homogeneous and known.  In practice, there
is no guarantee that the Hamiltonian of the system will be this
perfect. In particular the $J$-couplings between neighboring spins are
likely to be subject to random variations and the precise actual values
of the couplings may not be known, leading to an uncertainly in the
system Hamiltonian.  It is therefore very important to consider how
small random variations of the coupling strengths are likely to affect
the effectiveness of the control scheme.  To this end, let us assume
that the coupling between neighboring sites are random and vary around
an average value $J$, leading to an actual system Hamiltonian of the
form
\begin{equation}
\label{Ham_rand}
  H_s=J\bar H_0 = J\sum_{n=1}^{N-1} (1+\epsilon_n)
  \big(X_nX_{n+1}+Y_nY_{n+1}+ Z_n Z_{n+1}\big),
\end{equation}
where $\epsilon_n \in [-\alpha,+\alpha]$ is a uniformly distributed
random variable with mean $0$.  Assuming that the precise actual
couplings are not known, the best we can do is to study the average
effect of random variations of the coupling strengths of a certain
magnitude $\alpha$ on the entanglement generated.  To this end, we
evolve the ground state of the (randomly) perturbed Hamiltonian
(\ref{Ham_rand}) according to the Schr\"odinger equation with
$H=J\bar{H}_0+B(t)H_1$ with $B(t)$ being the pulse optimized for the
ideal Hamiltonian $H_0$ and its ground state $\ket{G_0}$.  The
resulting entanglement between the end spins will usually be less
than in the ideal case.  For comparison we generate 100 different
random Hamiltonians $\bar{H}_0$ for each value of $\alpha$, compute
the entanglement attained between the end spins as quantified by the
concurrence $C_{1N}$ if we apply the optimal pulse for the ideal
case, and take the average $\ave{C_{1N}}$.

\begin{figure}
\includegraphics[width=\columnwidth]{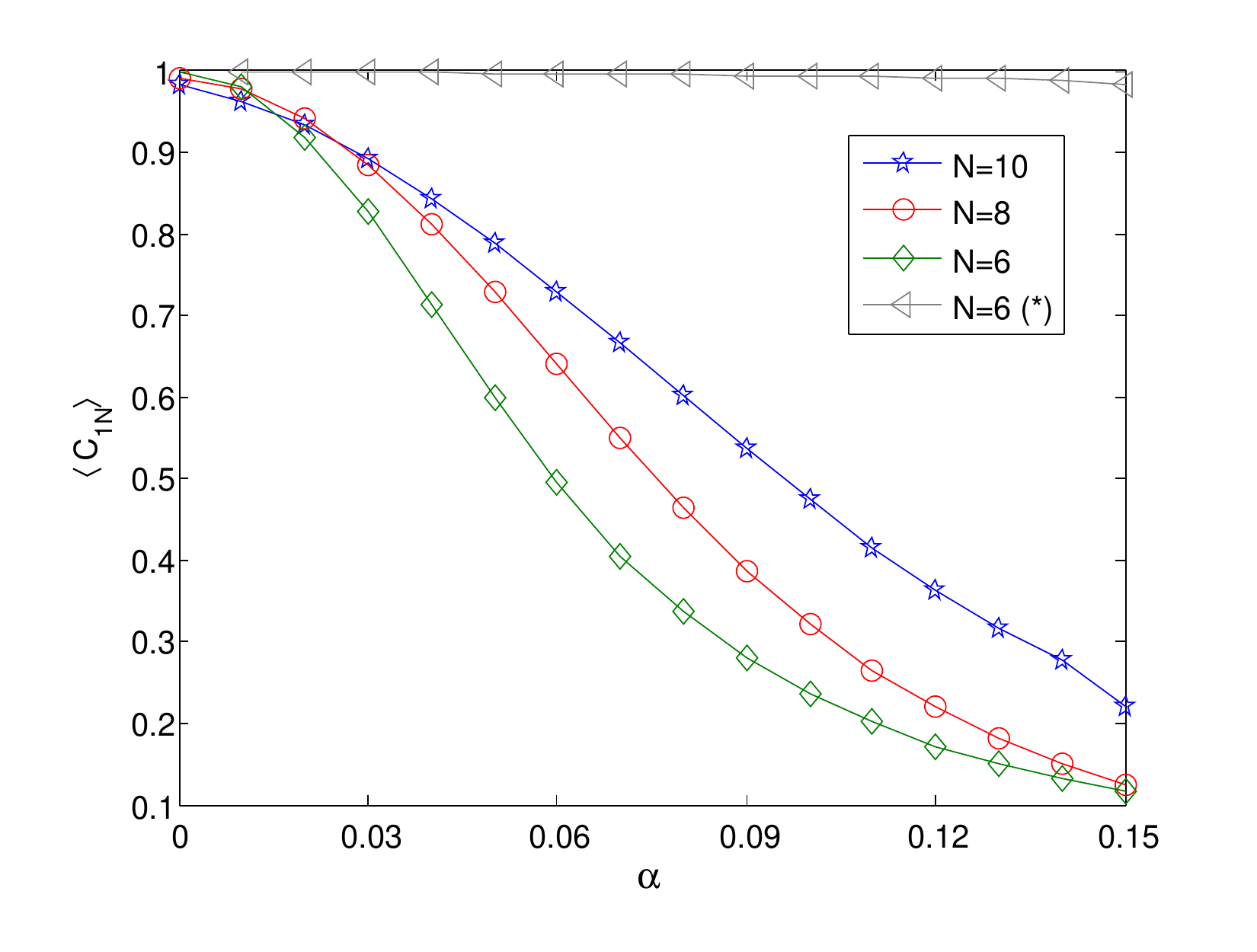} \caption{(Color online)
Average entanglement, $\ave{C_{1N}}$, in terms of disorder parameter
$\alpha$ for different lengths.  Each point is the average entanglement
over 100 different randomly perturbed Hamiltonians.  A small amount of
uncertainty in the coupling constants hardly affects the entanglement
achieved but for larger variations the entanglement drops significantly.
The average entanglement achieved for disordered systems when the actual
couplings are used in the optimization (solid black line with triangles
for $N=6$) is very close to one even for high disordered systems,
showing that inhomogeneity is not a limiting factor provided the actual
couplings can be estimated with sufficient accuracy.}  \label{Fig_rand}
\end{figure}

Fig.~\ref{Fig_rand} shows the average entanglement as a function of
$\alpha$.  Although the average entanglement decreases with
$\alpha$, it remains high for variations up to $\alpha=0.01J$,
showing that a small amount of uncertainty can be tolerated.  Larger
uncertainties in the coupling strength can be addressed by
experimental characterization using, e.g., spectral analysis or
Bayesian estimation techniques~\cite{SysIdent}, and the figure shows
that it is highly beneficial to experimentally characterize the
couplings to reasonable accuracy, or alternatively, to use adaptive
closed-loop experiments~\cite{PRA80n030301(R)} to refine the results
as disorder or inhomogeneity is not a limiting factor in the
optimization per se.  If the coupling constants used in the
optimization are sufficiently close to the actual couplings, near
perfect entanglement can be achieved even for inhomogeneous systems.
Another interesting feature of Fig.~\ref{Fig_rand} is the fact that
longer chains are \emph{less} sensitive to the random couplings, and
thus do not need to be characterized with the same accuracy
necessary for shorter chains to obtain the same amount of
entanglement.  This can be explained in terms of fundamental
mathematical properties and the \emph{central limit
theorem}~\cite{CLT}, which asserts that when $N$ identical random
variables are added, the distribution of the result is Gaussian and
its mean and standard deviation are normalized by $N$ and
$1/\sqrt{N}$ respectively.  This means that the randomness of the
system Hamiltonian (\ref{Ham_rand}) is effectively suppressed by a
factor of $\sqrt{N}$ with increasing $N$.

\subsection{Decoherence}

Another source of errors is from decoherence and population
relaxation induced by incoherent interactions of the system with the
environment. On the short time scales considered here we expect
decoherence due to pure dephasing to be the dominant effect and the
system evolution to be governed by a master equation of Lindblad
form
\begin{equation}
\label{eq:Lindblad} \dot{\rho}=-i[H(t),\rho]+ \L(\rho).
\end{equation}
To assess how such environmentally induced phase relaxation affects the
performance of the optimal control scheme, we compute the evolution of
the system according to (\ref{eq:Lindblad}) starting with the initial
state $\rho_0=\ket{G_0}\bra{G_0}$, subject to the Hamiltonian
$H(t)=JH_0+B(t)H_1$ as in (\ref{eqn:H}), where $B(t)$ is the optimal
pulse for the ideal system, and again calculate the concurrence between
the end spins of the final (mixed) state $\rho(t_f)$.

\begin{figure}
\includegraphics[width=\columnwidth]{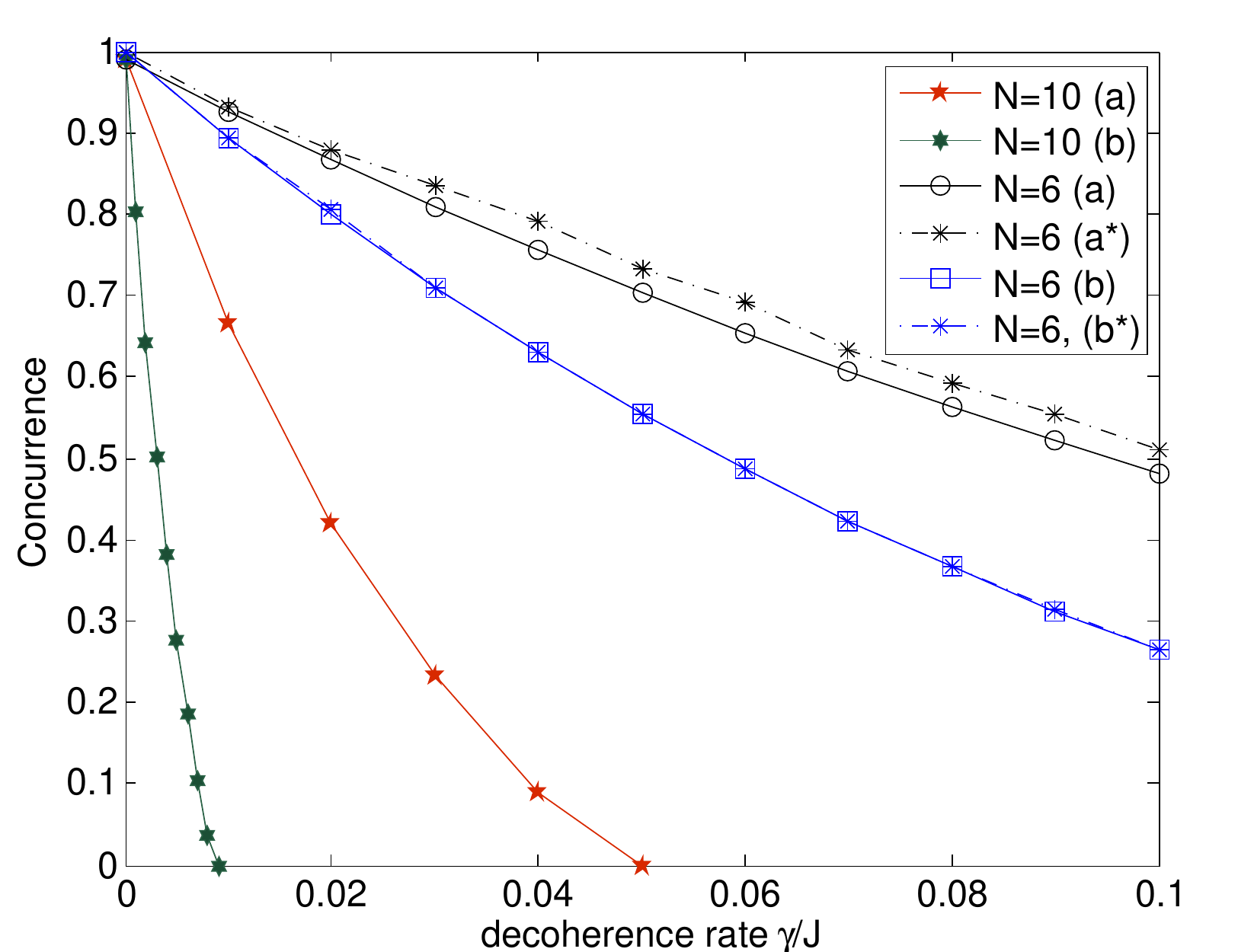}
\caption{(Color online) Entanglement as a function of dephasing rates
$\gamma/J$ for chains of length $N=6$ and $N=10$ for dephasing on the
end spins only (a) and dephasing of all qubits (b) as defined in
Eq.~(\ref{eqn:Lindblad2}) for a pulse optimized for the Hamiltonian
system, and for $N=6$ concurrences achieved with decoherence is taken
into account in the optimization ($*$).} \label{Fig_dec}
\end{figure}

Assuming that the main source of phase relaxation is from the
incoherent interactions with the control and measurement apparatus,
we consider two types of phase relaxation described by the Lindblad
operators
\begin{subequations}
  \label{eqn:Lindblad2}
    \begin{align}
    \L_1(\rho) &= -\gamma
    \{2\rho-Z_1\rho Z_1-Z_N\rho Z_N\}, \\
    \L_2(\rho) &= -\gamma
    \sum_{i=1}^N\{\rho-Z_i\rho Z_i\}.
   \end{align}
\end{subequations}
In the first scheme dephasing affects only the end spins, which are
interacting with the probes, while in the second scheme we assume
that all spins are affected by dephasing due to imperfect shielding.
The resulting entanglement between the end spins in terms of noise
strength $\gamma$ in both dephasing cases is shown in
Fig.~\ref{Fig_dec} for a chains of length $N=6$ and $N=10$.
Decoherence reduces the entanglement but when the noise strength is
small enough we are still able to generate entanglement with the
pulse optimized for the ground state.  When more spins are affected
by decoherence as in the second scenario, the entanglement decays
faster, and the effect of decoherence increases with chain length,
as expected.  For the scheme to be effective if all spins are
subject to dephasing for a chain of length $N=10$, the relative
decoherence rate $\gamma/J$ should be less than $0.1\%$, which is
consistent with usual quantum information processing requirements.
The results can generally be improved somewhat if decoherence is
taken into account in the optimization process but there are
limitations as coherent control cannot compensate for irreversible
effects caused by the Lindblad operators.  For $\gamma/J\ll 1$ we
find that the pulses obtained by optimization over the Lindblad
dynamics generate almost the same amount of entanglement as those
computed assuming Hamiltonian dynamics, and for longer chains
improvements seem to be even less although it must noted that there
are many solutions to the optimal control problem for Hamiltonian
systems that achieve near unit concurrence, and the robustness with
regard to decoherence of these pulses may vary.  Thus a more
extensive global search of the parameter space may yield better
results.

\section{Conclusion}

We have studied control of anti-ferromagnetic isotropic Heisenberg
chains using a single control field such as a magnetic field along a
given direction acting on a single site.  The intrinsic isotropic
Heisenberg Hamiltonian commutes with the total spin operator,
implying that the dynamics decomposes into $N+1$ independent
excitation subspaces. Although the control is not sufficient for
full controllability of the system as the resulting control
Hamiltonian commutes with the total spin operator, and hence each
excitation subspace is invariant under the dynamics regardless of
the control field applied, we verified that the system is
controllable in the largest excitation subspace, which guarantees
the existence of a control driving the ground state to a perfect
end-to-end entangled state in the subspace. Such control can be
generated by transforming the control problem to an optimization
problem and deriving the optimal control through standard
optimization procedure. More importantly, it was shown that the
resulting optimal control solutions are quite robust with regard to
multiple sources of imperfections, including imperfect initial state
preparation, incomplete confinement of the local control fields,
uncertainly about the couplings between adjacent spins and
decoherence due to environmental influences.

Furthermore, the quality of the entanglement achieved can be
substantially improved by taking the imperfections into account in the
optimization.  If we can estimate the amount of leakage of the control
field or the actual values of the couplings between adjacent spins using
experimental system identification techniques, then such information can
be incorporated in the optimization to generate greatly improved optimal
pulses.  Furthermore, we found that a control pulse that maximizes the
entanglement between the end spins starting with a high-temperature
thermal ensemble is automatically optimal for lower-temperature initial
ensembles, which means that the optimal control scheme is effectively
independent of the initial state, provided it is a thermal ensemble.
Another interesting observation is that simulations strongly suggest
that for our problem controllability on the largest excitation subspace
holds not only for the isotropic Heisenberg chain but also for a general
XXZ model, which deserves future investigation.

\section{Acknowledgements}

We gratefully thank Pierre de Fouquieres, Peter Pemberton-Ross and
Alastair Kay for valuable discussions. SGS acknowledges funding from
EPSRC ARF Grant EP/D07192X/1, the QIPIRC and Hitachi. XW thanks the
Cambridge Overseas Trust, Hughes Hall and the Cambridge Philosophical
Society for support.  AB and SB are supported by the EPSRC.  SB is also
supported by the QIPIRC (GR/S82176 /01), the Royal Society and the
Wolfson Foundation.


\end{document}